\documentclass[conference]{IEEEtran}
\IEEEoverridecommandlockouts
\usepackage{cite}
\usepackage{amsmath,amssymb,amsfonts}
\usepackage{algorithmic}
\usepackage{graphicx}
\usepackage{textcomp}
\usepackage{multirow}
\usepackage{xcolor}
\usepackage{makecell}
\usepackage{enumitem}
\usepackage{placeins}
\usepackage{hyperref}
\usepackage{dblfloatfix}
\usepackage{hyperref}
\usepackage{graphics}
\usepackage{tcolorbox}
\usepackage{xcolor}
\usepackage{subcaption}
\usepackage[english]{babel}
\usepackage [autostyle, english = american]{csquotes}
\MakeOuterQuote{"}
\def\BibTeX{{\rm B\kern-.05em{\sc i\kern-.025em b}\kern-.08em
    T\kern-.1667em\lower.7ex\hbox{E}\kern-.125emX}}
\definecolor{skyblue}{RGB}{135, 206, 235}
\begin{document}

\title{Requirements Elicitation Follow-Up Question Generation}

\author{
    \IEEEauthorblockN{Yuchen Shen, Anmol Singhal, Travis Breaux}
    \IEEEauthorblockA{Carnegie Mellon University, Pittsburgh, USA\\ Email: \{yuchens2, singhal2, tdbreaux\}@andrew.cmu.edu}
}

\maketitle

\begin{abstract}
Interviews are a widely used technique in eliciting requirements to gather stakeholder needs, preferences, and expectations for a software system. Effective interviewing requires skilled interviewers to formulate appropriate interview questions in real time while facing multiple challenges, including lack of familiarity with the domain, excessive cognitive load, and information overload that hinders how humans process stakeholders' speech. Recently, large language models (LLMs) have exhibited state-of-the-art performance in multiple natural language processing tasks, including text summarization and entailment. To support interviewers, we investigate the application of GPT-4o to generate follow-up interview questions during requirements elicitation by building on a framework of common interviewer mistake types. In addition, we describe methods to generate questions based on interviewee speech. We report a controlled experiment to evaluate LLM-generated and human-authored questions with minimal guidance, and a second controlled experiment to evaluate the LLM-generated questions when generation is guided by interviewer mistake types. Our findings demonstrate that, for both experiments, the LLM-generated questions are no worse than the human-authored questions with respect to clarity, relevancy, and informativeness. In addition, LLM-generated questions outperform human-authored questions when guided by common mistakes types. This highlights the potential of using LLMs to help interviewers improve the quality and ease of requirements elicitation interviews in real time.

\end{abstract}


% \maketitle
\vspace{-8pt}
\section{Introduction}
\label{section:intro}
% Introduce requirements elicitation & interviews
In requirements engineering, business analysts employ interviews to elicit, acquire, identify, and elaborate requirements for information systems~\cite{ZC05}. Interviews allow analysts to directly engage with stakeholders who describe their experiences and points of view, clarify ambiguity, and provide context about their specific needs. By asking well-formulated questions, business analysts can probe topics in more depth and uncover requirements that may not otherwise emerge. However, variations in domain knowledge, communication skills, and cultural differences can all lead to different outcomes in elicitation. Interviewers may find it stressful to listen to an interviewee's speech while identifying appropriate follow-up questions during the interview, because of psychological hindrances such as excessive cognitive load. Good follow-up questions must be clear, relevant to the interviewee's speech, and informative to draw out tacit knowledge and hidden needs that may otherwise be overlooked. Interviews are also costly to conduct, require training and significant time and effort to plan, schedule, conduct, and analyze responses~\cite{AKO20, CDJ14}, which makes it difficult to scale interviews to large groups of stakeholders.

Since interviews involve natural language dialogue, both from interviewee speech and interviewer questions, this has made interviews a suitable beneficiary of recent advances in natural language processing. In particular, autoregressively trained large language models (LLMs) such as generative pre-trained transformers (GPTs) have been used to process and generate natural language for open-ended tasks where the structure of the output is not predetermined, including interview script generation~\cite{GA23}, user simulation~\cite{ACG+24}, transcript analysis~\cite{SJB+24}, and chatbots for information elicitation~\cite{HZT+21}. These techniques focus primarily on the preparation and post-processing of interviews, with relatively limited research studying elicitation at interview time. 

In this paper, we report on the effectiveness of LLMs in generating follow-up interview questions. We describe the following contributions: 1) an empirical analysis of 14 interviews to identify the quantity of prior context needed for an interviewer to formulate a question; 2) experimental results comparing minimally guided GPT-4o-generated and human-authored follow-up questions for relevancy, clarity, and informativeness; 3) a framework synthesized from prior work to outline common mistakes made by interviewers during elicitation; and 4) experimental results comparing GPT-4o-generated and human-authored follow-up questions to address common interviewer mistakes. The findings show that in general, GPT-4o-generated questions are no worse or better than human-generated questions. However, when focusing on specific interviewer mistakes, GPT-4o-generated questions are rated more highly for relevancy, clarity, and informativeness. 

The remainder of this paper is organized as follows: in Section~\ref{section:background}, we present background and related work; in Section~\ref{section:approach} we present our approach, including the experimental designs; in Section~\ref{section:results}, we report our evaluation results; in Section~\ref{section:discussion}, we discuss our findings; in Section~\ref{section:threats}, we discuss threats to validity, and we conclude in Section~\ref{section:conclusion}.

\section{Background and Related Work}
\label{section:background}

We now review background and related work on requirements elicitation via interviews.

\subsection{Requirements Elicitation via Interviews}
\label{subsection:eliciation}

% Describe interview elicitation as a method of preference elicitation, different forms of interviews
Interviews are one of the most traditional and commonly used requirements elicitation techniques, providing a systematic and interactive way for requirements analysts to directly engage with stakeholders and ask questions to understand their needs, preferences, and expectations about a product or service~\cite{HD03, PFQ21, SFB18, ZC05}. During interviews, interviewers can delve deep into the interview domain and ask questions to elicit stakeholder preferences, resolve possible ambiguities around multifaceted, hard to express, or even conflicting viewpoints~\cite{ZC05}, and gain insight into tacit knowledge the interviewees might harbor~\cite{FSG16}, ultimately obtaining a list of requirements co-created by the participating requirements analyst and stakeholder~\cite{FSD22}. Interviews may be conducted in a one-to-one format, in which an interviewer directly interviews a stakeholder, or within a group, in which multiple stakeholders and analysts can participate~\cite{SS14}. Interviews may be structured, where a predetermined list of questions is prepared and asked during the interview; unstructured, with no preparation of a script or questions in advance; or semi-structured, where interviewers adapt their questions to the conversation based on interviewee responses during the interview time~\cite{KPJ+16, ZC05}. Structured interviews are more rigorous, but can limit the investigation of new ideas, while unstructured interviews may have the risk of neglecting important topics~\cite{ZC05}.

% Describe challenges
Despite the variety of interview formats, conducting interviews involves numerous challenges. Stakeholder preferences are often tacit knowledge that requires significant effort to extract~\cite{FSG16}. In particular, knowledge experts may become more skilled but less aware of the cognitive process involved. Moreover, due to the limited storage capacity of attentional memory, the knowledge generated is limited to that readily available to the conscious introspection of the expert at the time of inquiry~\cite{AT90}. Interviewers need specific skills, such as interpersonal skills to build rapport with interviewees~\cite{BZF+18}, and sufficient domain knowledge and subject matter expertise on the interview topic, to ask relevant, meaningful and thought-provoking questions~\cite{BZF+18}, understand interviewee responses~\cite{FSG16}, and identify gaps and opportunities to follow up and delve deeper for further elicitation~\cite{HSK14}. Cultural and linguistic barriers can also hinder communication~\cite{FSD22, ZC05}. Interviewers should consciously avoid tunneling, which is when interviews talk excessively about one topic and do not cover the necessary breadth of interview topics~\cite{Duhigg16}. Interviewers also need to suppress jumping between different topics, which can result in less cohesive experiences and more superficial requirements~\cite{BWS18}. Psychologically, since interviews are conducted in real time, interviewers may need to manage the stress associated with developing appropriate follow-up questions. Follow-up questions are important because they can clarify ambiguity, validate interviewee statements, and explore an unexpected topic in more detail~\cite{Barton15}. This stress may arise from an excessive cognitive load due to having to communicate back and forth with interviewees and, at the same time, process interviewee responses~\cite{Hanway20}, or from information overload, where interviewees provide long or complicated responses that may contain multiple points of interest that are difficult to conceptually organize for an interviewer less experienced in the domain~\cite{ASD12}. Misinterpretations and cognitive limitations during communication can hinder the use of interviews for elicitation.

% Describe the mistake criteria
The existing literature has described various criteria for practitioners in the appropriate conduct of elicitation interviews, including criteria for the appropriate opening and closing of interviews, the right atmosphere and flow, the question framing, the question content, the proper elicitation goals, and avoidance of common mistakes made during elicitation interviews. For example, interviewers may incorrectly open or close an interview by asking for ideas without providing appropriate context or guidelines~\cite{DFS+17, BZF+18}, failing to build rapport before asking questions~\cite{BZF+18}, or not providing a summary before the interview ends~\cite{DFS+17, BZF+18}. Ambiguities may not be properly leveraged, including tacit assumptions~\cite{PTA94} or tacit knowledge known to stakeholders but not to the interviewer~\cite{DFS+17, BZF+18}, or they may not be properly elicited, unclear or contradictory speech made by stakeholders may not be properly addressed~\cite{DFS+17, BZF+18}, and alternatives may not be properly considered~\cite{PTA94, BZF+18}. Questions may be too generic or domain independent, or have been misunderstood~\cite{BZF+18}. Implicit stakeholder goals may not be explicitly stated~\cite{DFS+17}, and implicit stakeholders may not be identified~\cite{BZF+18, BZF+18}. Feature priority~\cite{BZF+18} and resource limits~\cite{DFS+17} may not be clearly elicited. Interviewers may go back and forth between topics that lead to confusion, or run out of questions due to an interrogatory interview style~\cite{DFS+17}. Questions may not be properly framed, such as questions that are too long or overly articulated~\cite{DFS+17, BZF+18}, too vague, technical, or irrelevant~\cite{BZF+18}, inappropriate to the stakeholder profile~\cite{BZF+18}, involve multiple kinds of requirements~\cite{DFS+17}, ask for solutions~\cite{BZF+18}, and contain jargon~\cite{BZF+18}, among other weaknesses. Finally, interviews should not be unethical or disrespectful~\cite{HZT+21}. A framework of common interviewer mistakes based on past literature is shown in Section~\ref{section:mistake_study}.

\subsection{Automated Techniques}

% Describe automatic techniques that help ease interviews, including prev. work on pre-interview and post-interview aids, introduce LLM and its role and potential
Natural Language Processing (NLP) techniques provide many opportunities to automate the elicitation process. Advances in textual entailment have been transformative in the generation of high-quality human language~\cite{AM10, DRZ+22}. Recognizing textual entailment (RTE), which is a class of techniques to test whether a hypothesis $h$ follows from a premise $p$, have been extended in RTE2 to question-answering (Q\&A) tasks that introduce a question $q$, such that $p$:$q \implies h$, where $h$ is the answer to the question over the content of the premise $p$~\cite{Pol20}. Matsumoto et al. (2018) further proposed reformulating the Q\&A task to predict the question $q$ from the premise and hypothesis. Whereas Matsumoto et al. generate questions that can be answered from a premise, we are interested in generating questions that cannot be answered from the premise but that directionally derive from the premise to yield new information when posed to an interviewee. In particular, autoregressive training techniques for building large language models (LLMs) such as Generative Pre-trained Transformers (GPTs) have improved the ability to understand and generate natural language texts~\cite{BMR+20}. Whereas more traditional state-of-the-art models like BERT are bidirectional transformers trained with separate tasks (a masked language modeling task and a next sentence prediction task) in order to teach them to understand natural language syntactically and semantically, autoregressive models like GPTs are large-scale models, unidirectionally trained to predict each upcoming word in the text given all previous words, which leads to improved ability to generate coherent and contextually appropriate text as opposed to just natural language understanding. Due to the improvements of LLMs on natural language generation, the latter provides more flexibility for open-ended tasks where the length and structure of the output are not predetermined. LLMs have been explored and shown to be successful in various requirements engineering tasks, such as use case model extraction~\cite{BGG+22} and goal modeling~\cite{CCH+23,SB25}, among others. Interviews, as an important requirement artifact, is an example of an open-ended task that may benefit from LLMs. For example, LLMs are used to generate simulated users for interviews and analysis to elicit requirements~\cite{ACG+24}; to generate requirements elicitation interview scripts~\cite{GA23, GA24}; and to post-process interview transcripts~\cite{SJB+24}. Moreover, some of the work attempted at building chatbots that are effective at engaging users and eliciting quality information, but found challenges such as the lack of clarity for generated interview questions, and chatbot-unrecognized user input~\cite{HZT+21}.

% Describe how our work may differ from previous work

Prior research focuses mainly on preparing and post-processing interviews, whereas research to assist the elicitation process during the interview is limited. In this paper, we report efforts to bridge the gap using an LLM-assisted method to guide the generation of follow-up questions to support requirements elicitation.

\section{Approach}
\label{section:approach}

We investigate LLM-based ways to generate follow-up questions for interviewers using the preceding interview context to entail a good follow-up question. To that end, we introduce the following research questions (RQs):

\textbf{RQ1}: What are the types of follow-up questions in an interview and how many prior speaker turns are needed to formulate each question? 

\textbf{RQ2}: How do minimally guided LLM-generated follow-up questions compare to human-generated questions?

\textbf{RQ3}: To what extent can an LLM decide whether a human-generated question demonstrates a common interviewer mistake type?

\textbf{RQ4}: When provided with common interviewer mistake types to avoid, how do LLM-generated questions compare to human-generated questions?

% \textbf{RQ5}: How well do LLM-generated questions simultaneously avoid multiple interviewer mistake types?

To answer these questions, we first conducted a literature review to identify prior work related to interview follow-up questions and specifically drawn from research on interviewer mistakes. Next, we designed and conducted three experiments: (1) an experiment to measure quality-related differences between questions generated by GPT-4o compared to questions raised by trained interviewers within the same interview context; and (2) an experiment to assess how well GPT-4o can detect whether interviewer-raised questions demonstrate any one of a number of common interviewer mistakes; and (3) an experiment to measure the quality-related difference between GPT-4o generated questions and human analyst-authored questions, when both the GPT-4o and the human are instructed to avoid a common mistake. Below, we outline the data collection process for obtaining interviewer transcripts, followed by a detailed discussion of the experimental designs and evaluation methods.

\subsection{Interview Transcript Collection}
\label{subsection:transcripts}
We collected interview transcripts by hiring and training interviewers to interview individuals about their experiences using directory services, which are web and mobile applications designed to help users find specific products or services (e.g., apartment finding apps). We hired four interviewers who are students enrolled in the [blinded degree program] who had at least two years of industrial software engineering experience in a full-time software engineering position. Two years is approximately the number of years needed to reach Software Development Engineer II, which is a non-entry-level position. In addition, the interviewers had completed two courses in requirements engineering and product management, covering topics that include identifying customer value and modeling and analyzing requirements. Each interviewer was trained by the investigators on requirements elicitation using Ferrari et al.'s course~\cite{FSB+19}, entitled ``Learning Requirements Elicitation Interviews with Role-Playing, Self-assessment, and Peer-review.'' Finally, all interviewers completed training on the protocols for protecting human subjects in accordance with the Institutional Review Board (IRB), which monitored the data collection process. 

To schedule interviews, the hired interviewers provided a total of fourteen 20-minute time slots. Interviewees were recruited by email. To qualify, interviewees must be at least 18 years old, have experience using web and mobile applications, and be fluent in English. Before recruitment, interviewees must sign an informed consent form regarding their rights and protection of personal data. Participants are compensated with a \$25 Amazon Gift Card after completing the interview. 

Each interview was randomly assigned to one of four directory service domains: apartment finding, restaurant finding, hiking trail finding, and clinic finding. The interviewers were provided with a general question upon which to base their interviews, corresponding to the assigned domain, as follows:

\textbf{Apartment}: How do you find an apartment?

\textbf{Restaurant}: How do you choose a restaurant to eat at?
  
\textbf{Hiking}: How do you plan a trail hike in a park?

\textbf{Health Clinic}: How do you choose a clinic to visit when you get sick?

The main focus of these interviews is to elicit stakeholder preferences, which are a class of non-functional requirements that describe qualities that a stakeholder desires in a software system~\cite{SB24}. All interviews were conducted online through Zoom with the video turned on and the interview automatically recorded. After each interview, the recording transcript was retrieved and anonymized before further analysis. 

\subsection{Study 1: Minimally Guided Questions}
\label{subsection:comparative_study}

The \textbf{RQ1} is answered by the second author manually reviewing the collected transcripts to identify all interviewer questions and to determine the minimum number of preceding conversational turns upon which a question depends. A question depends on a prior turn if the context provides the information needed to comprehend the question. We define a \textbf{turn} as a single person's speech in the conversation from start to finish. For example, if an interviewer poses a question and the interviewee responds, this exchange constitutes two turns. Our analysis reports the turn frequency, which is the number of turns required to establish a sufficient context for an interviewer question. This analysis also produced a dataset of 146 follow-up question contexts, mapping the preceding relevant turns to the corresponding interviewer questions. In the paper, we refer to this data set as the \textit{ turn context data set}.

In addition to turn frequency, the second author used open coding~\cite{Sal12} to classify the types of follow-up questions in the turn context dataset. To conduct this analysis, the author first assigned a label to each question in the dataset based on their understanding of the relationship between the question and the prior dependent turns. For example, when an interviewer asks about two topics A and B in turn $n-2$, followed by the interviewee responding to only topic A in turn $n-1$, followed by the interviewer asking the interviewee to respond to topic B, the second follow-up question is labeled `question probing' and its dependent on two prior turns. For each question in the dataset, the author labeled the questions by inspecting the prior turns, reusing labels when appropriate, and maintaining definitions for the labels. The coding process saturated when the second author had coded 32 questions, after which the author discovered no new labels while coding the remaining questions. This process yielded a total of seven labels that we presented in Section~\ref{section:results}.

We answer RQ2 and evaluate the GPT-4o-generated questions against the human-authored questions by relying on Paul Grice's pragmatic theory of productive conversations that includes four maxims: be relevant; be clear; be informative; and be truthful. While truthfulness is important in requirements elicitation, particularly when interviewing an adversarial stakeholder, in this research, we chose to assume that stakeholders are truthful. Thus, we adopt the remaining three maxims to evaluate questions, summarized, and presented to the raters:

\begin{itemize}
    \item \textit{relevancy} - an interviewer question should be relevant to the topic or system and be situated within the flow of conversation
    \item \textit{clarity} - an interviewer question should be clear and understandable by the stakeholder
    \item \textit{informativeness} - an interviewer question should result in a stakeholder response that increases the quantity of information known about the system
\end{itemize}

Formally, we evaluate the following falsifiable hypothesis using an independent t-test:

\textbf{H1:} Minimally-guided LLM-generated questions are not different from human-generated questions with respect to relevancy, clarity, and informativeness. 

RQ2 and H1 were answered and tested, respectively, by asking human's to judge the difference between GPT-4o-generated and human-authored questions. First, we randomly sampled 20 questions from the turn context dataset. Next, we observed that the sampled questions and their prior turns, called an \textit{instance}, contained speech artifacts and grammatical errors, likely due to audio transcription errors. To reduce the influence of grammar on human perception of question quality, we manually corrected the grammar of each sampled instance using Grammarly \footnote{https://app.grammarly.com/} before passing the instance as input to GPT-4o.

For each instance, we prompted GPT-4o to generate a follow-up question by only providing the prior turns and by omitting the interviewer's follow-up question. We prompt GPT-4o-2024-08-06, a state-of-the-art, closed-source, multi-modal model developed by OpenAI. The following prompt template shown in Figure~\ref{fig:prompt1} was used to generate the follow-up questions:

\begin{figure}[ht]
\centering
\begin{subfigure}[t]{\linewidth}
\centering
\begin{tabular}{p{8.5cm}}
\fontsize{7}{5}{\texttt{You are an AI agent capable of generating context summaries. During a requirements elicitation interview with an interviewee about how the interviewee conducts \{interview domain\}, the INTERVIEWEE and INTERVIEWER have had the following conversation: \{interview turns\}. Generate a follow-up question that the INTERVIEWER should ask next based on the conversation. Restrict your response to only show the follow-up question without explanation.}}
\end{tabular}
\caption{Prompt 1 to Generate Minimally Guided Questions}
\vspace{10pt}
\label{fig:prompt1}
\end{subfigure}

\begin{subfigure}[t]{\linewidth}
\centering
\begin{tabular}{p{8.5cm}}
\fontsize{7}{5}{\texttt{You are an AI agent capable of conducting requirements elicitation interviews. During a requirements elicitation interview with an interviewee about how the interviewee conducts \{domain keyword\}, the INTERVIEWEE said '\{interviewee speech\}'. Then the INTERVIEWER asked a follow up question by saying '\{interviewer question\}'. Standard: \{mistake criterion\}. Please classify based solely on whether the INTERVIEWER’s response meets this specific standard, and refrain from using any other standards related to follow up questions when you classify. If the INTERVIEWER’s response meets this standard, output 'Yes', otherwise output 'No'. Restrict your response to output only 'Yes' or 'No' without explanations.}}
\end{tabular}
\caption{Prompt 2 to Classify Follow-up Questions}
\vspace{10pt}
\label{fig:prompt2}
\end{subfigure}

\begin{subfigure}[t]{\linewidth}
\centering
\begin{tabular}{p{8.5cm}}
\fontsize{7}{5}{\texttt{You are an AI agent capable of conducting requirements elicitation interviews. During a requirements elicitation interview with an interviewee about how the interviewee conducts \{domain keyword\}, the INTERVIEWEE said '\{interviewee speech\}'. Generate a follow-up question that meets the following criterion based ONLY on what the INTERVIEWEE said, and restrict your response to only show the follow-up question without explanation. Criterion: \{mistake criterion\}}}
\end{tabular}
\caption{Prompt 3 to Generate Mistake-Guided Questions}
% \vspace{-12pt}
\vspace{10pt}
\label{fig:prompt3}
\end{subfigure}

\caption{LLM Prompts Used in the Studies}
\label{fig:prompts}
% \vspace{-12pt}
\end{figure}

% \begin{figure}[ht]
% \begin{tabular}{p{8.5cm}}
% \fontsize{7}{5}{\texttt{You are an AI agent capable of generating context summaries. During a requirements elicitation interview with an interviewee about how the interviewee conducts \{interview domain\}, the INTERVIEWEE and INTERVIEWER have had the following conversation: \{interview turns\}. Generate a follow-up question that the INTERVIEWER should ask next based on the conversation. Restrict your response to only show the follow-up question without explanation.}}
% \end{tabular}
% \caption{Prompt 1 to Generate Minimally Guided Questions}
% \label{fig:prompt1}
% \end{figure}

% \texttt{You are an AI agent capable of generating context summaries. During a requirements elicitation interview with an interviewee about how the interviewee conducts \{interview domain\}, the INTERVIEWEE and INTERVIEWER have had the following conversation: \{interview turns\}. Generate a follow-up question that the INTERVIEWER should ask next based on the conversation. Restrict your response to only show the follow-up question without explanation.}

Since follow-up question generation is a creative task, we used the model's default temperature setting of \textbf{1.0} to promote non-deterministic outputs, which means the model could generate different paraphrases of the same question or even different questions from the same prompt.

This experiment consists of two groups of survey respondents: one of the respondents who see only the GPT-4o-generated questions and one of the respondents who see only human-authored questions. We designed two surveys per group with 10 questions per survey to reduce the overall time taken to complete the survey, while increasing the number of questions evaluated by each group to 20 questions. The instructions for each survey were the same. 

Each survey consisted of ten question blocks that were randomized for each participant to randomly distribute any ordering effects among questions. A question block consists of a one-sentence description of the domain, the transcript context preceding the question, the question, followed by three semantic scales for relevancy, clarity, and informativeness as shown below, respectively:
\begin{itemize}
    \item Please rate the question's relevance with respect to the interview conversation.
    \item Please rate the question's clarity.
    \item Please rate the question's ability to encourage the interviewee to provide an informative response.
\end{itemize}
Respondents choose a level on a 6-point scale from least to greatest, e.g., very irrelevant, irrelevant, somewhat irrelevant, somewhat relevant, relevant, very relevant. The other two scales were similarly labeled, replacing irrelevant/relevant with unclear/clear and uninformative/informative.

The surveys were published using the Qualtrics platform~\footnote{https://www.qualtrics.com/}. We recruited 32 survey participants to rate the questions in each survey (64 participants in each group) using the Prolific platform~\footnote{https://www.prolific.com/}. To enroll in the experiment, participants first signed an informed consent form, which included information about their rights and confidentiality protection. In addition, the participants must be at least 18 years old, reside in the US at the time of the study, and speak English as their primary language. Participants were assigned to either the GPT-4o-generated or the human-authored question group, and participants were not informed about which group they were assigned to. Upon completion of the survey, each participant was paid \$6 as compensation through Prolific.

The distribution of the survey results were tested for normality using the Shapiro-Wilk normality test. Since normality held, the results were then used to perform a two-tailed Student's T-test to test hypothesis H1, which we report in Section~\ref{section:results}.

\subsection{Synthesized Mistake Framework}
\label{section:mistake_study}

We answer RQ3 and RQ4 by investigating the source of interviewer mistakes during interviews and by synthesizing a set of mistake criteria to inform how to generate a follow-up question. We first review the method used to synthesize these criteria before describing two studies to answer the RQs.

Requirements elicitation research has considered the criteria that support formulating and asking good interview questions. We identified 14 published papers by keyword searching Google Scholar\footnote{https://scholar.google.com/} in multiple fields, including requirements engineering, human-computer interaction, and software engineering. A paper was selected if it contains criteria that describe either a) criteria that a good interview question should satisfy; or b) mistake criteria that an interview question should avoid. A criterion is identified when a statement describing what good interview questions should do, or what should be avoided, is made in a paper. When extracting criteria from the papers, if a criterion mentioned describes a good interview question without describing the mistake counterpart, we manually change the description to the mistake counterpart. For example, a good question should elicit only one kind of requirement each time~\cite{DFS+17}, which can be restated as a mistake criterion to ``ask a question that involves multiple kinds of requirements.'' This process yielded 28 mistake criteria covering a range of concerns, including leading with the wrong interview opening, failing to construct the right interview ambiance, the wrong follow-up questions, the wrong way to frame a question, poor question flow, failing to meet elicitation goals, and wrong closing of an interview. 

From the 28 criteria\footnote{The list of 14 referenced papers and the 28 interview mistake criteria is provided in the repository https://github.com/anmolsinghal98/Requirements-Elicitation-Follow-Up-Question-Generation}, we down-selected to focus on 14 criteria that describe mistakes in two categories: 1) \textit{follow-up questions}, which are questions asked as a follow-up to stakeholder's prior speech; and 2) \textit{question framing}, which refers to how a question should be framed or formulated during the interview. We selected these two categories to focus our question generation efforts on because a large majority of requirements elicitation interview questions depend on stakeholders' prior speech in order to be formulated, and all questions need proper formulation, which means good follow-up questions and correct question framing are essential components to a successful elicitation interview.

The mistake criteria for the selected categories are as follows. Each criterion is followed by one or more citations of papers from which the criterion is derived. 

\textbf{Follow-up questions:}
\begin{itemize}
    \item \textit{Fail to elicit tacit assumptions}, i.e., fail to justify or authorize assumptions stakeholders tacitly make without justification.~\cite{PTA94}
    \item \textit{Fail to consider alternatives}, i.e., fail to look for alternative information or alternatives to existing requirements.~\cite{PTA94, BZF+18} 
    \item \textit{No clarification when unclear}, i.e., accepts what interviewee said without asking for clarification when words interviewee said are unclear.~\cite{DFS+17, SJB+24}
    \item \textit{No clarification when contradictory}, i.e., accepts what interviewee said without asking for clarification when words interviewee said are contradictory.~\cite{DFS+17, SJB+24}
    \item \textit{Fail to elicit tacit knowledge}, i.e., fail to elicit tacit knowledge known to interviewee but unknown to interviewer.~\cite{AT90, DFS+17, BZF+18, ACG+24, SJB+24}
\end{itemize}

\textbf{Question framing: }
\begin{itemize}
    \item \textit{Ask a generic, domain-independent question}, i.e.,  fail to ask a question related to the interview domain or the question asked is too generic.~\cite{BZF+18, GA23}
    \item \textit{Ask a question that is too long or articulated}, i.e., ask a question too long or complicated that would likely require interviewee to ask for repeating or rephrasing multiple times.~\cite{FSB+19, DFS+17, BZF+18, GA23}
    \item \textit{Use jargon}, i.e., ask a question containing special words or expressions not in the common vocabulary and is difficult for interviewees to understand.~\cite{BZF+18, GA24}
    \item \textit{Ask a technical question}, i.e., ask a question that requires technical knowledge in order to answer.~\cite{FSB+19, BZF+18, GA23}
    \item \textit{Ask a question inappropriate to user's profile}, i.e., ask a question that cannot be answered by the interviewee given the interviewee's profile.~\cite{BZF+18}
    \item \textit{Ask for solutions}, i.e., ask interviewee to present a solution to satisfy a requirement.~\cite{BZF+18, GA23}
    \item \textit{Ask a question that involves multiple kinds of requirements}, i.e., mix different categories of requirements or multiple specific requirements within one category into a single question.~\cite{DFS+17}
    \item \textit{Ask a vague question that leads to multiple interpretations}, i.e., ask a question that can be interpreted in more than one way.~\cite{FSB+19, HZT+21, BZF+18, GA23, GA24}
    \item \textit{Ask a vague question which could infer no reasonable meaning}, i.e., ask a question that does not have enough context or clarity for interviewee to answer.~\cite{FSB+19, HZT+21, BZF+18, GA23, GA24}
\end{itemize}

We study mistake-guided question generation in two studies: first, a human analyst who was familiar with the mistake criteria but unaware of the method and results for generating follow-on questions was asked to classify whether the human-authored question demonstrates one of the 14 interviewer mistake types above, and, if it does, the human analyst is next asked to write a suitable question that avoids the mistake. In this study, we prompt GPT-4o to perform the same classification and question generation task on the same data set, and then perform a comparative evaluation of the GPT-4o and human analyst results. Second, we collect questions for which both GPT-4o and the human analyst determine that the interviewer question demonstrates one of the fourteen mistake types and survey an online population to compare the questions. 

\subsection{Study 2: Mistake-Guided Question Classification}
\label{subsection:study2_classification}

We answer RQ3 by comparing the classification results of GPT-4o and the human analyst. For this study, we use GPT-4o-2024-08-06 with temperature set to 1.0. The human analyst in this study has three years of industry experience working as a RE researcher, and the analyst was excluded from the research steps to prompt the model and review the model output to avoid introducing bias into this evaluation. We now describe the details of our method.

With the transcripts collected in Section~\ref{subsection:transcripts}, the first author segmented the transcript into interviewee-interviewer pairs for each domain. Next, this author labeled the pairs where the interviewer speech included a question, and removed all remaining pairs. Finally, this author randomly sampled 30 pairs and asked the analyst to separately classify whether the interviewer question demonstrates the fourteen mistake criteria shown in Section~\ref{section:mistake_study}. For 30 questions pairs, this yields $30 \times 14 = 420$ classifications. To prompt GPT-4o to perform the same task, the first author designed a prompt template to be used once for each question-pair and criterion, which is shown in Figure~\ref{fig:prompt2}:

% \begin{figure}[ht]
% \begin{tabular}{p{8.5cm}}
% \fontsize{7}{5}{\texttt{You are an AI agent capable of conducting requirements elicitation interviews. During a requirements elicitation interview with an interviewee about how the interviewee conducts \{domain keyword\}, the INTERVIEWEE said '\{interviewee speech\}'. Then the INTERVIEWER asked a follow up question by saying '\{interviewer question\}'. Standard: \{mistake criterion\}. Please classify based solely on whether the INTERVIEWER’s response meets this specific standard, and refrain from using any other standards related to follow up questions when you classify. If the INTERVIEWER’s response meets this standard, output 'Yes', otherwise output 'No'. Restrict your response to output only 'Yes' or 'No' without explanations.}}
% \end{tabular}
% \caption{Prompt 2 to Classify Follow-up Questions}
% \label{fig:prompt2}
% \end{figure}

The \texttt{\{domain keyword\}} is one of "apartment", "restaurant", "trail", "clinic". In addition, when designing the prompts for mistake criteria ``No clarification when contradictory,'' ``Ask technical questions,'' ``Ask questions inappropriate to user's profile,'' and ``Ask for solutions,'' we add a one-shot example to illustrate the meaning of the criterion in order to improve performance. We discuss insights from the prompt tuning process in Section~\ref{subsection: pro_des}.

Whenever the human analyst identified that a question demonstrates a mistake type, the human analyst is asked to proceed to generate a suitable question that avoids the mistake. To fully complement the human analyst data, the first author designed a second prompt to generate questions that avoid the mistake type identified by GPT-4o. The following prompt template shown in Figure~\ref{fig:prompt3} was used to collect these questions.

% \begin{figure}[ht]
% \begin{tabular}{p{8.5cm}}
% \fontsize{7}{5}{\texttt{You are an AI agent capable of conducting requirements elicitation interviews. During a requirements elicitation interview with an interviewee about how the interviewee conducts \{domain keyword\}, the INTERVIEWEE said '\{interviewee speech\}'. Generate a follow-up question that meets the following criterion based ONLY on what the INTERVIEWEE said, and restrict your response to only show the follow-up question without explanation. Criterion: \{mistake criterion\}}}
% \end{tabular}
% \caption{Prompt 3 to Generate Mistake-Guided Questions}
% \vspace{-12pt}
% \label{fig:prompt3}
% \end{figure}

To answer RQ3, we compare the GPT-4o classification results with the human analyst classification results for the 420 classification instances. Ahmed et al. (2025) argue that LLM-based labeling of software engineering artifacts is approaching human-level performance for some classification tasks, including code summarization, causality, and semantic similarity~\cite{ADT+25}. In these tasks, differences of human opinion are measured using inter-rater reliability. Building on this prior work, the RQ3 aims to assess the relative performance of LLM-based classification by calculating the agreement rate, which is the percentage of instances where the GPT-4o classification agrees with that of the human analyst, along with specific details of how the agreement rate is distributed across different mistake types, and report the results in Section~\ref{subsection:mis_typ_clas_res}.

\subsection{Study 3: Mistake-Guided Question Generation}
\label{subsection:mis_qgen}

After collecting all GPT-4o-generated and human analyst-authored questions, we constructed a second dataset that consisted of the 30 records containing the interviewee speech, interviewer question, domain keyword, and mistake criterion identified by the analyst. Next, we duplicated this data into two sets: one with the interviewer question replaced by the GPT-4o-generated question, and one with the interviewer question replaced by the human analyst question for the given criterion. This yielded a dataset consisting of 128 total records, which exceeds the number of records required by a power analysis with medium effect size 0.5, power 0.8, and significance level 0.05. Next, we randomly shuffled the combined dataset, placed the 128 records into 32 surveys, wherein each survey describes four follow-up questions.

The 32 surveys each contained four pairs of a GPT-4o generated question and one human analyst authored question, including the same interviewee speech, domain keyword, and mistake criterion assigned to the pair. The source of the questions (GPT-4o or human analyst) was not indicated and both pairs of questions and the order of the two questions in each pair were randomized. Participants were asked to choose which of the two questions better avoids the mistake explained in the mistake criterion. In addition, participants were asked to evaluate the relevancy, clarity, and informativeness of the questions using a 5-point semantic scale from least to greatest, e.g., very irrelevant, somewhat irrelevant, neutral, somewhat relevant, very relevant. 

The survey was published using Qualtrics, and we recruited 32 participant from the Prolific platform, where each participant was randomly assigned to one of the surveys. Prior to enrolling in the study, each participant must complete the informed consent form describing their rights and confidentiality protection. In addition, participants must be at least 18 years old, currently reside in the United States, and speak English as their primary language. Upon completion of the survey, each participant is paid \$6 as compensation through Prolific.

We test the following falsifiable hypothesis:

\textbf{H2}: When provided with 14 types of interviewer mistakes to avoid, LLM-generated questions are not different from human-generated questions with respect to avoidance of mistakes, relevancy, clarity, and informativeness.

We evaluated the ability to avoid the mistake types using a mixed-effect Bradley-Terry model, which are preferred in paired comparison experiments. The Bradley-Terry model is a generalization of a binomial model that can be used to predict the preference between two items in a paired comparison~\cite{BT52}. We chose to use the mixed-effect Bradley-Terry model instead of the standard model because each survey taker in our survey study rated four question pairs, which means the responses are not independent. A mixed-effect Bradley-Terry model accounts for this dependency by incorporating random effects for each rater, allowing us to model individual differences in rating tendencies, and thus mitigates bias introduced by variance in individual perceptions and ensures a more reliable estimation of the relative quality of the questions. Further, the mixed-effect model allows us to generalize findings beyond the specific raters in our study, making the results more robust and reflective of the actual question quality. 

We evaluate relevancy, clarity, and informativeness scores provided by survey respondents using a mixed-effect ordinal logistic regression model. This model deals with ordinal data and allows us to have fixed effect for the question source, and random effects for each rater, similar to the mixed-effect Bradley-Terry model.

\section{Results}
\label{section:results}
We now report our results for each study described in Section~\ref{section:approach}.

\subsection{Minimally Guided Question Generation Results}
\label{subsection:comparative_results}

\begin{figure}[ht]
\centering
\begin{subfigure}[t]{\linewidth}
    \centering
    \includegraphics[scale=.5]{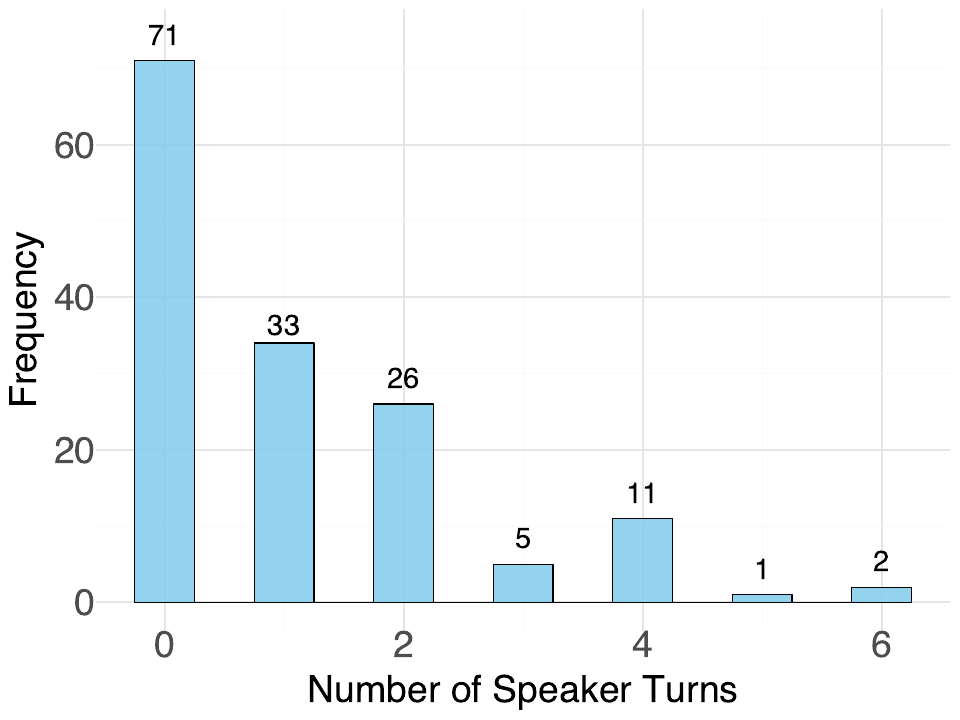}
    \caption{Speaker Turn Distribution Required By Follow-up Questions}
    \label{fig:speaker_turn_dist}
    \vspace{10pt}
\end{subfigure}

\begin{subfigure}[t]{\linewidth}
    \centering
    \includegraphics[scale=.5]{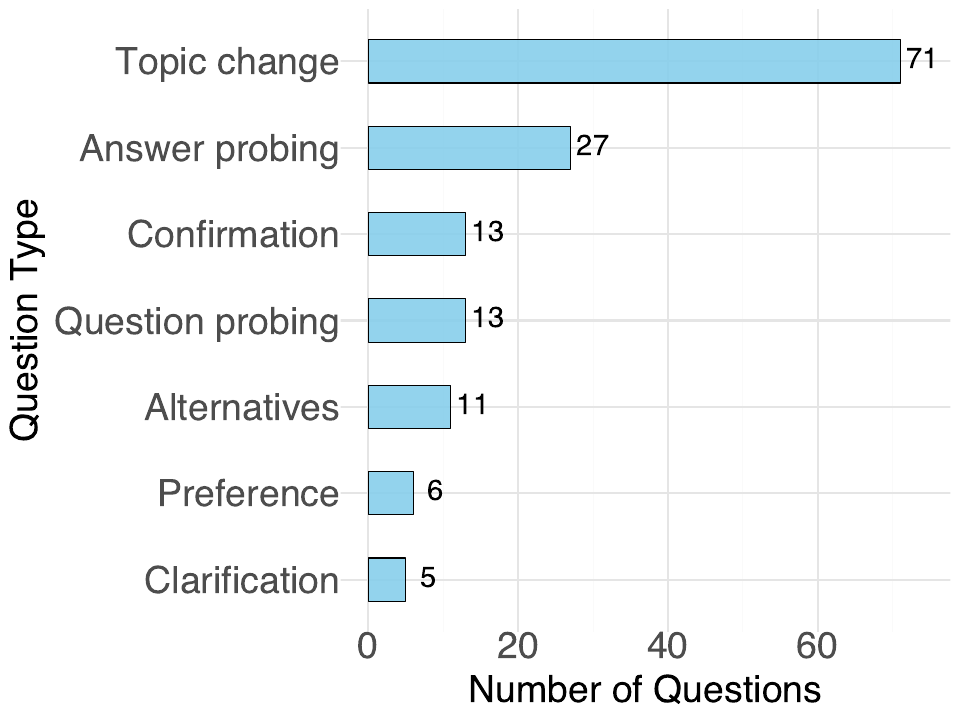}
    \caption{Follow-up Question Type Distribution}
    \label{fig:question_type_dist}
    \vspace{10pt}
\end{subfigure}
\caption{Speaker Turn and Follow-up Question Type Distributions}
\label{fig:dists}
\end{figure}

In Section~\ref{subsection:comparative_study}, RQ1 asks ``What are the types of follow-up questions in an interview and how many prior speaker turns are needed to formulate each question?'' Figure \ref{fig:speaker_turn_dist} reports the speaker turn distribution for number of turns required by each follow-up question. On an average, the turn context size needed to formulate a question is one, i.e., providing the interviewee's last response was sufficient to formulate a follow-up question. We also note that 71/146 follow-up questions require no context, 104/146 require \textit{up to one speaker turn}, and 98\% (144/146) of questions require \textit{up to four turns}. 

% \begin{figure}[ht]
%     \centering
%     \includegraphics[scale=.4]{speaker_turns_distribution.pdf}
%     \caption{Speaker Turn Distribution Required By Follow-up Questions}
%     \label{fig:speaker_turn_dist}
% \end{figure}

% \begin{figure}[ht]
%     \centering
%     \includegraphics[scale=.4]{question_type_distribution.pdf}
%     \caption{Follow-up Question Type Distribution}
%     \label{fig:question_type_dist}
% \end{figure}

During our analysis of the required speaker context, we open-coded the interviewer's follow-up questions to infer the interviewer's possible motivation behind raising the question. The following seven-question typology was derived from the context and aims to describe this possible motivation:

\begin{itemize}
    \item \textbf{Topic change}: Questions in which the interviewer changes the conversation topic to talk about a different and unrelated topic than the one represented by the prior turns. 
    \item \textbf{Answer probing}: Questions in which the interviewer appears to further probe a concept that the interviewee mentioned in their last turn. 
    \item \textbf{Confirmation}: Questions in which the interviewer repeats or paraphrases the interviewee response to ensure that they correctly understand the interviewee's statement.
    \item \textbf{Question probing}: Questions in which the interviewer asks about a concept within the current topic scope but that was missing from the interviewee's last turn(s). 
    \item \textbf{Alternative-seeking}: Questions in which the interviewer broadly asks about alternatives to a concept being discussed (e.g., what else questions).
    \item \textbf{Preference-seeking}: Questions in which the interviewer expects a yes/no answer from the interviewee to accept or reject an otherwise provisional interviewee preference. 
    \item \textbf{Clarification}: Questions in which the interviewer asks about an ambiguous or vague concept in the interviewee's last turn. 
\end{itemize}

Figure \ref{fig:question_type_dist} presents the follow-up question typology distribution. When we review this distribution by question type, we observed that the number of turns needed as context varies. Topic change questions require no prior speaker turns, whereas answer probing, confirmation, and clarification questions are primarily formulated using one speaker turn. Question probing, alternative-seeking, and preference-seeking questions require at least two prior turns.

%\begin{tcolorbox}[colback=skyblue!20, colframe=skyblue!70!black, title=Answer to RQ1]
%Interviewers may use up to seven types of follow-up questions. On an average, the turn context size needed to formulate a question is one turn. A context size of four prior speaker turns covers 98\% of follow-up questions in the dataset. 
%\end{tcolorbox}

The RQ2 asks ``How do minimally guided LLM-generated follow-up questions compare to human-generated questions?'' To assess performance differences between GPT-4o-generated and human-authored questions, we conducted a two-tailed Student's T-test for relevancy, clarity, and informativeness (see Table~\ref{table:match_res1}). The results show no statistically significant differences between the two groups for any metric for $p \le 0.05$. 

\begin{table}[ht]
\centering
\caption{GPT-4o vs Human Study Results}
\label{table:combined_results}
\renewcommand{\arraystretch}{1.2}
\begin{subtable}[t]{\linewidth}
\centering
\caption{Study 1 Results: Human vs GPT-4o}
\resizebox{\columnwidth}{!}{%
\begin{tabular}{ l c c c } 
 \hline
  \textbf{Metric} & \textbf{Informativeness} & \textbf{Relevancy} & \textbf{Clarity} \\ 
  \hline
 Human Avg Score & 4.6 & 4.8 & 4.9 \\
 GPT-4o Avg Score & 4.8 & 5.0 & 5.1 \\
  p-value & 0.17 & 0.08 & 0.10 \\
  \hline
\end{tabular}%
}
\label{table:match_res1}
\end{subtable}

\vspace{1cm}

\renewcommand{\arraystretch}{1.6}
\begin{subtable}[t]{\linewidth}
\centering
\caption{Classification Results For Each Mistake Type}
\resizebox{\columnwidth}{!}{%
\begin{tabular}{ l c c c c } 
 \hline
  % \textbf{Mistake Type} & \textbf{Human} & \textbf{GPT-4o} & \textbf{Agreement Rate} & \textbf{\textcolor{red}{Avoidance Rate}} \\ 
    \textbf{Mistake} & 
    \textbf{Hu-} & 
    \textbf{GPT-} & 
    \textbf{Agreement} & \textbf{Avoidance} \\ \textbf{Type} & \textbf{man} & \textbf{4o} & \textbf{Rate} & \textbf{Rate} \\ 
  \hline
 \makecell[l]{Fail to elicit tacit assumptions} & 19 & 25 & 66.7\% & 76.7\% \\ 
 Fail to consider alternatives & 30 & 28 & 93.3\% & 26.7\% \\ 
No clarification when unclear & 20 & 16 & 60.0\% & 70.0\%  \\ 
 No clarification when contradictory & 7 & 2 & 83.3\% & 83.3\% \\ 
 Fail to elicit tacit knowledge & 21 & 20 & 70.0\% & 90.0\% \\ 
\makecell[l]{Ask a generic, \\domain-independent question} & 7 & 10 & 90.0\% & 96.7\% \\ 
 Ask a question that is too long \\ or articulated & 11 & 7 & 86.7\% & 90.0\% \\ 
 Use jargon & 7 & 1 & 80.0\% & 100.0\% \\ 
 Ask a technical question & 5 & 1 & 86.7\% & 100.0\% \\ 
 \makecell[l]{Ask a question inappropriate to \\ user's profile} & 5 & 0 & 83.3\% & 100.0\% \\ 
 Ask for solutions & 6 & 2 & 86.7\% & 100.0\% \\ 
 \makecell[l]{Ask a questions that involves \\ multiple kinds of requirements} & 7 & 12 & 76.7\% & 60.0\% \\ 
 \makecell[l]{Ask a vague question that leads to \\ multiple interpretations} & 27 & 27 & 93.3\% & 93.3\% \\ 
 \makecell[l]{Ask a vague question which could \\ infer no reasonable meaning} & 5 & 12 & 76.7\% & 93.3\%  \\ 
 \makecell[l]{Total} & 177 & 163 & 81.0\% & 84.3\% \\ 
\hline
\end{tabular}%
}
\label{table:class_res}
\end{subtable}

\vspace{1cm}

\begin{subtable}[t]{\linewidth}
\centering
\caption{Relevancy, Clarity and Informativeness Results}
\resizebox{\columnwidth}{!}{%
\begin{tabular}{ l c c c } 
 \hline
  \textbf{Metric} & \textbf{Relevancy} & \textbf{Clarity} & \textbf{Informativeness} \\ 
  \hline
 Human Avg Score & 3.5 & 3.9 & 3.6 \\
 GPT-4o Avg Score & 4.4 & 4.5 & 4.1 \\
 Human Win Rate & 21.1\% & 17.2\%  & 25.8\%  \\
  GPT-4o Win Rate & 59.4\% & 42.2\% & 47.7\% \\
  Tie Rate & 19.5\% & 40.6\% & 26.6\% \\
  p-value & $1.21 \times 10^{-10}$ & $1.34 \times 10^{-6}$ & $1.77 \times 10^{-5}$ \\
  \hline
\end{tabular}%
}
\vspace{6pt}
\label{table:match_res}
\end{subtable}

\end{table}

% \begin{center}
% \begin{table}[ht]
% \centering
%  \renewcommand{\arraystretch}{1.2}
% \vspace{-12pt}
% \caption{Study 1 Results: Human vs GPT-4o}
% % \vspace{8pt}
% \resizebox{\columnwidth}{!}{%
% \begin{tabular}{ l c c c } 
%  \hline
%   \textbf{Metric} & \textbf{Informativeness} & \textbf{Relevancy} & \textbf{Clarity} \\ 
%   \hline
%  Human Avg Score & 4.6 & 4.8 & 4.9 \\
%  GPT-4o Avg Score & 4.8 & 5.0 & 5.1 \\
%   p-value & 0.17 & 0.08 & 0.10 \\
%   \hline
% \end{tabular}%
% }
% \vspace{-12pt}
% \label{table:match_res1}
% \end{table}
% \end{center}
% \FloatBarrier

\subsection{Mistake Types Classification Results}
\label{subsection:mis_typ_clas_res}

In Section~\ref{subsection:study2_classification}, we described our study to answer RQ3, which asks ``To what extent can an LLM decide whether a human-generated question demonstrates a common interviewer mistake type?''

We answer RQ3 by comparing GPT-4o and human classifications of whether an interviewer question demonstrates one of 14 mistake types. A total of 420 classification were separately obtained from GPT-4o and a human analyst. We observe that GPT-4o and the human analyst agree 81.0\% (340/420) of the time. Specifically, the human analyst classified 177/420 while GPT-4o classified 163/420 instances as having demonstrated a given mistake type. The classification distribution is presented in the first four columns of Table~\ref{table:class_res}, where the first column is the mistake type, the second and third columns are the human's and GPT-4o's classification count for demonstrating the mistake type, respectively, and the last column is the agreement rate.

% \begin{center}
% \begin{table}[ht]
% \centering
%  \renewcommand{\arraystretch}{1.2}
% \vspace{-12pt}
% \caption{Classification Results For Each Mistake Type}
% % \vspace{8pt}
% \resizebox{\columnwidth}{!}{%
% \begin{tabular}{ l c c c } 
%  \hline
%   \textbf{Mistake Type} & \textbf{Human} & \textbf{GPT-4o} & \textbf{Agreement Rate} \\ 
%   \hline
%  \makecell[l]{Fail to elicit tacit assumptions} & 19 & 25 & 66.7\% \\ 
%  Fail to consider alternatives & 30 & 28 & 93.3\% \\ 
% No clarification when unclear & 20 & 16 & 60.0\% \\ 
%  No clarification when contradictory & 7 & 2 & 83.3\% \\ 
%  Fail to elicit tacit knowledge & 21 & 20 & 70.0\% \\ 
% \makecell[l]{Ask generic, domain-independent \\ questions} & 7 & 10 & 90.0\% \\ 
%  Ask questions too long or articulated & 11 & 7 & 86.7\% \\ 
%  Use jargon & 7 & 1 & 80.0\% \\ 
%  Ask technical questions & 5 & 1 & 86.7\%\\ 
%  \makecell[l]{Ask questions inappropriate to \\ user’s profile} & 5 & 0 & 83.3\% \\ 
%  Ask for solutions & 6 & 2 & 86.7\% \\ 
%  \makecell[l]{Ask questions that involve \\ multiple kinds of requirements} & 7 & 12 & 76.7\% \\ 
%  \makecell[l]{Ask vague questions that lead to \\ multiple interpretations} & 27 & 27 & 93.3\% \\ 
%  \makecell[l]{Ask vague questions which could \\ infer no reasonable
% meaning} & 5 & 12 & 76.7\% \\ 
%  \makecell[l]{Total} & 177 & 163 & 81.0\% \\ 
% \hline
% \end{tabular}%
% }
% \vspace{-12pt}
% \label{table:class_res}
% \end{table}
% \end{center}
% % \FloatBarrier

Finally, we observed 128 questions where both GPT-4o and the human analyst classified a given mistake type. These instances are further analyzed and described in Section~\ref{subsection:mis_qgen}.

\subsection{Mistake-guided Question Generation Results}
\label{subsection:mis_generation_res}

In Section~\ref{subsection:mis_qgen}, RQ4 asks ``When provided with common interviewer mistake types to avoid, how do LLM-generated questions compare to human-generated questions?'' To answer RQ4, survey respondents were presented with one turn of interviewee speech, with a question pair consisting of one interviewer question generated by GPT-4o and one authored by the human analyst, and a mistake type. The respondent was asked to select the question that best avoids the mistake type. Out of the 128 questions pairs, the GPT-4o-generated questions were selected 87 times, while the human questions were selected 41 times, indicating respondents believed that GPT-4o was more proficient at addressing the mistake type in 68.0\% of question pairs. 

The mixed-effect Bradley-Terry model reported a p-value $4.23 \times 10^{-8} \le 0.05$, in which case we reject the null hypothesis \textbf{H2} that there is no difference between GPT-4o and the human analyst. The probability that GPT-4o generates a better question that avoids a given mistake type is about $2.662 / (1 + 2.662) = 93.5\%$ based on the odds ratio of 2.662.

Survey respondents were asked to rate on an ordinal scale of 1 to 5 for how relevant, clear, and informative each question is. A mixed-effect ordinal logistic regression model was used to analyze the scores, which is reported in Table~\ref{table:match_res}, including the GPT-4o and the human average score, question win rate for relevancy, clarity, informativeness, the tie rate, and the p-value.

% \begin{center}
% \begin{table}[ht]
% \centering
%  \renewcommand{\arraystretch}{1.2}
% \vspace{-12pt}
% \caption{Relevancy, Clarity and Informativeness Results}
% % \vspace{8pt}
% \resizebox{\columnwidth}{!}{%
% \begin{tabular}{ l c c c } 
%  \hline
%   \textbf{Metric} & \textbf{Relevancy} & \textbf{Clarity} & \textbf{Informativeness} \\ 
%   \hline
%  Human Avg Score & 3.5 & 3.9 & 3.6 \\
%  GPT-4o Avg Score & 4.4 & 4.5 & 4.1 \\
%  Human Win Rate & 21.1\% & 17.2\%  & 25.8\%  \\
%   GPT-4o Win Rate & 59.4\% & 42.2\% & 47.7\% \\
%   Tie Rate & 19.5\% & 40.6\% & 26.6\% \\
%   p-value & $1.21 \times 10^{-10}$ & $1.34 \times 10^{-6}$ & $1.77 \times 10^{-5}$ \\
%   \hline
% \end{tabular}%
% }
% \vspace{-12pt}
% \label{table:match_res}
% \end{table}
% \end{center}
% % \FloatBarrier

Among the 128 question pairs, GPT-4o scored better on average than the human-authored questions with respect to all three quality criteria, and the p-values are all less than 0.05. 

%With our results, we answer RQ4 as follows:

%\begin{tcolorbox}[colback=skyblue!20, colframe=skyblue!70!black, title=Answer to RQ4]
%When evaluated on the question's ability to avoid a given mistake type, and on the relevancy, clarity, and informativeness, LLM-generated questions are better than human analyst-authored questions. Our statistical tests provided strong evidence to support this conclusion.
%\end{tcolorbox}

\section{Discussion}
\label{section:discussion}
We now discuss our prompt design insights, context in question generation, question quality, and anecdotal results from a side study of GPT-4o's efficacy to generate follow-up questions that simultaneously avoid multiple mistake types.

\subsection{Prompts Design}
\label{subsection: pro_des}
We designed the GPT-4o prompts using persona-based prompting technique~\cite{WFH+23}, consisting of the system prompt ``You are an AI agent capable of conducting requirements elicitation interviews.''

When designing the prompts, our tuning and experimentation process revealed that GPT-4o failed to distinguish between the interviewer and interviewee speech. This challenge manifests itself as a classification error in which the model incorrectly attributes speech or misinterprets the intent of the question. We found that a simple typographical modification by capitalizing the role identifiers "INTERVIEWER" and "INTERVIEWEE" significantly reduced the classification error. This finding is consistent with previous research, which reports that LLM performance is sensitive to separators and capitalization changes~\cite{SCT+23}.

In addition, because LLM performance is shown to be poor in the presence of negation~\cite{KS20}, we reformulated the mistake description from a negative to a positive framing. For example, we included the instruction that ``a good follow-up question should consider alternatives'' instead of ``a good follow-up question should not fail to consider alternatives.'' We found that 4/14 mistake types produced notably lower performance, for these we added a one-shot demonstration as recommended by prior work~\cite{BMR+20}. To illustrate, we added ``For example, it's inappropriate to ask users about how to design a specific feature, or what would an ideal user interface look like.'' to clarify the meaning of the ``ask for solutions'' mistake type. 
For 9/14 mistake types in the classification prompts, we added a step-by-step instruction. For example, in the ``no clarification when contradictory'' mistake type, we added ``To classify whether the INTERVIEWER's question meets this standard, first consider if the INTERVIEWEE mentioned anything contradictory. If it does not, then the standard is met. Otherwise, look at whether the INTERVIEWER's question attempts to clarify the contradiction.''
\vspace{-4pt}
\subsection{Context in Question Generation}
\label{con_for_que_gen}

In our analysis of the number of turns or context required to ask a follow-up question, we observed that 70\% of follow-up questions require zero or one prior speaker turns, which suggests that interviewers rely primarily on the most recent speech when formulating questions. This observation indicates that interviewers may lack capacity to generate questions from longer context windows due to psychological processes, such as cognitive load. This has two consequences: (1) LLM-supported tools may effectively generate human-comparable follow-up questions by focusing on the immediate context, reducing the need for extensive conversational history. In fact, a context window of four prior speaker turns appears sufficient to account for 98\% of all the follow-up questions observed. In addition, (2) LLM-supported tools could provide above-human performance if they can enable human interviewers to track information across larger context windows when generating follow-up questions.

The typology of the follow-up questions has practical implications. For instance, recognizing that topic change questions typically require no prior context could help systems initiate new conversational directions efficiently. Meanwhile, understanding that answer probing, confirmation, and clarification questions are closely tied to the most recent turn could guide models in prioritizing recent content when formulating these question types, whereas for question probing, alternative-seeking, and preference-seeking questions that demand a longer context.

\subsection{Question Quality}
\label{que_qua}

In this work, we conducted a minimally guided question study that shows that GPT-4o achieves comparable performance to human interviewers when generating follow-up questions. The mistake-guided question study shows that GPT-4o generates better questions than a human analyst with regard to relevancy, clarity, and informativeness. The improvement may be attributed to how the mistake types naturally align with the quality criteria of relevancy, clarity, and informativeness. For example, avoiding the mistake ``fail to consider alternatives,'' may yield questions with better informativeness, or avoiding the three mistakes ``ask questions too long or articulated,'' ``ask vague questions that lead to multiple interpretations,'' and ``ask vague questions which could infer no reasonable meaning'' may yield better clarity. Finally, avoiding the two mistakes ``Ask generic, domain-independent questions,'' and ``ask questions inappropriate to user's profile'' may yield more relevant questions. By guiding the LLM to avoid these mistake types, the LLM may generate questions that are more performant on the evaluation criteria. If true, then research to better understand the underlying causes behind failure to perform elicitation in other ways, i.e., the underlying mistakes, could improve LLM instructional design to generate better interview questions.

\subsection{Side Study: Simultaneous Avoidance of Mistakes}
In Study 3, we investigated how well an LLM can generate questions to avoid a single mistake type. We conducted a side study to preview how well LLM-generated questions can simultaneously avoid multiple interviewer mistake types. This research question is interesting because simultaneous avoidance could yield questions of even better quality. However, prior work shows that LLMs are less performant when multiple constraints must be conjointly satisfied in a single task~\cite{GHM+23, VOS+22}. When prompting GPT-4o to simultaneously avoid all fourteen interviewer mistake types listed in Section~\ref{section:mistake_study}, and to evaluate whether a GPT-4o-generated question successfully avoids each mistake type using the same 30 interviewee-interviewer pairs from Study 3, we observed that 66/420 classifications still demonstrate a given mistake type. The last column in Table~\ref{table:class_res} presents the success avoidance rate for each mistake type. In addition, 1/30 questions successfully avoided all mistake types, and 28/30 questions successfully avoided at least 11 mistake types. This shows that while an LLM may avoid a majority of mistake types simultaneously, more work is needed to evaluate and improve the avoidance rate. However, we did not evaluate how these generated questions compare with human-authored questions using the same Study 3 protocol. We postpone that question to future work.

% \begin{center}
% \begin{table}[ht]
% \centering
%  \renewcommand{\arraystretch}{0.9}
% \vspace{-12pt}
% \small
% \caption{Classification Results For Each Mistake Type}
% % \vspace{8pt}
% \resizebox{\columnwidth}{!}{%
% \begin{tabular}{ l c c c } 
%  \hline
%   \textbf{Mistake Type} & \textbf{Avoidance Rate} \\ 
%   \hline
%  \makecell[l]{Fail to elicit tacit assumptions} & 76.7\% \\ 
%  Fail to consider alternatives & 26.7\%  \\ 
% No clarification when unclear & 70.0\%  \\ 
%  No clarification when contradictory & 83.3\%  \\ 
%  Fail to elicit tacit knowledge & 90.0\%  \\ 
% \makecell[l]{Ask generic, domain-independent \\ questions} & 96.7\% \\ 
%  Ask questions too long or articulated & 90.0\%  \\ 
%  Use jargon & 100.0\%  \\ 
%  Ask technical questions & 100.0\% \\ 
%  \makecell[l]{Ask questions inappropriate to \\ user’s profile} & 100.0\% \\ 
%  Ask for solutions & 100.0\% \\ 
%  \makecell[l]{Ask questions that involve \\ multiple kinds of requirements} & 60.0\%  \\ 
%  \makecell[l]{Ask vague questions that lead to \\ multiple interpretations} & 93.3\%  \\ 
%  \makecell[l]{Ask vague questions which could \\ infer no reasonable
% meaning} & 93.3\% \\ 
%   \hline
% \end{tabular}%
% }
% \vspace{-14pt}
% \label{table:reclass_res}
% \end{table}
% \end{center}
% % \FloatBarrier

\section{Threats to Validity}
\label{section:threats}
\textit{Construct validity} refers to whether we are measuring what we believe we are measuring~\cite{Yin09}. In our research, the interview contexts used were obtained from real interviews conducted by trained interviewers. The questions in Study 1 were obtained from these transcripts. We reused the contexts in Studies 2 and 3. However, the questions were authored by a human analyst who was not engaged in the interview and who had access to the mistake catalog during question formulation. In addition, the quality of the question is difficult to measure. To address this threat, we adopted Paul Grice's pragmatic theory of productive conversations, specifically relevancy, clarity, and informativeness, to measure quality. Moreover, the interviewer mistake types were synthesized from literature review, which speaks to their broad observation across multiple studies, but which may not comprehensively cover all possible mistakes.

\textit{Internal validity} refers to validity of the analyses and conclusions drawn from the data~\cite{Yin09}. In our evaluation, the survey participants' pre-existing biases about AI-generated content could influenced their judgments. To reduce this threat, we randomly shuffled questions and blinded raters to the source of the questions (human or GPT-4o). For example, in Study 3, the question source was blinded, and the order of the questions was randomized to further evenly distribute any impact of human raters being biased toward the first or last shown question. The human analyst who classified interviewer mistakes and generated alternative questions may also have pre-existing biases about AI-generated content. We mitigated this threat by preventing the human analyst from learning about the LLM generation procedures and by concealing the survey designs and hypotheses from the analyst. In addition, humans may encounter cognitive challenges, such as fatigue and attentional limitations, when tasked with classifying interviewer mistakes and generating questions that avoid the mistakes as part of the research. This may produce questions perceived as less relevant, clear, or informative than if a human performed the same task without cognitive challenges. In contrast, LLMs are prone to hallucination, i.e., generating fluent but incorrect or misleading content~\cite{FKK+24}. This means that some GPT-4o-generated questions may appear professionally crafted or linguistically advanced, while omitting necessary domain-specific information, or they do not adequately advance the elicitation objectives. Our evaluation, which relies on specific quality assessments, does not specifically detect hallucination. However, if hallucinations correlate with quality decrement, then the quality measurements would be impacted. In addition, the second author performed the open coding and analysis of the interview transcripts to answer RQ1 without formal validation measures. Systematic bias in the question categorization may exist, which can affect the reliability of the results. 

\textit{External validity} refers to the generalizability of the results~\cite{Yin09}. Our work mainly focused on requirements elicitation interviews in four directory service domains and on eliciting stakeholder preferences. Therefore, the generalization of our results to other types of requirements elicitation interviews is unknown. Although internet users spend large amounts of time searching for information online, and thus observations from interviews about directory services could generalize to other Internet applications, these interviews did not cover subjects with safety-critical or performance requirements, such as autonomous driving or medical device software. In addition, preference elicitation is a specific variant of interviewing that may not be representative of all types of interview. Hence, our analysis that relied on the collected interview transcripts may not adequately generalize to all types of requirements elicitation interviews. 

The interview transcripts were created by interviewers with at least two years of software engineering experience who were trained to conduct interviews using Ferrari et al's techniques~\cite{FSB+19}. Although these interviewers were professionals returning for graduate study, the interviews were conducted in an academic setting. Therefore, these results may not extend to industrial settings. In addition, while the interviewees were required to have domain expertise, the interviewers were not required to have this expertise and were not assigned the job role to interview users for an app they were developing. These differences mean that human performance observed in response to RQ2 may not reflect the capabilities of expert practitioners, thus overstating the relative performance of LLM-generated questions. Finally, while the interviewees are all users of apps in the category for which they were recruited to be interviewed, they may not be representative of the population of users for any specific app.

The evaluators of the LLM- and human-generated questions were recruited using the Prolific platform. Thus, the quality measurements of the question reflect the opinions of internet users, and they were not individuals trained in discourse analysis. As a result, the evaluators may interpret the evaluation criteria, i.e., relevance, clarity, and informativeness, differently from trained analysts when rating or comparing follow-up questions, and thus the average ratings could be different from those analysts. According to Douglas et al., in comparison to the US population, the Prolific population is 67.5\% female, skews younger, and and consists of people that are similar in age and income~\cite{DEB23}. We limited respondents to a US location and native English speakers.

We used GPT-4o as LLM in our studies and the results may not be generalized to other language models. The rapid pace of LLM development means that newer models may demonstrate behaviors different from those we observed here. The prompts we used for our task may also not transfer to other models.

Our approach studied how well GPT-4o generates follow-up interview questions that traditionally require significant human expertise. However, we neither assumed nor demonstrated that GPT-4o can perform the full range of interviewing capabilities described in Section~\ref{subsection:eliciation}, such as building rapport with interviewees or adapting to cultural differences. Our evaluations focused narrowly on question generation quality rather than comprehensive interviewing competence. In addition, we did not empirically study whether our methods could address the cost and scaling challenges of traditional elicitation interviews.
	
\section{Conclusion and Future Work}
\label{section:conclusion}
In this paper, we proposed a framework that describes common interviewer mistake types and uses to evaluate the quality of follow-up questions. We collected interview data and conducted studies to test the capability of GPT-4o in classifying whether a question demonstrates a mistake type, and to evaluate GPT-4o-generated questions against human-authored questions in controlled experiments. Our findings demonstrate that minimally guided GPT-4o-generated questions are no better or worse than human-authored questions with respect to clarity, relevancy, and informativeness. However, LLM-generated questions outperformed human-authored questions in mistake-guided question generation. We believe our findings indicate that GPT-4o can be employed to support requirements elicitation interviews to enhance human performance.

Although we did not fully incorporate the question generation process into interviews, we believe that it is possible to do so by collecting interview transcripts in real-time and presenting up to four speaker turns to the model with prompts aimed at generating mistake-avoiding questions. Presenting these questions to the interviewer for their review and possible selection could yield more efficient interviews. In future work, we envision studying this performance in the context of variable communication skills and cultural differences, as well as excessive cognitive load.

Other sophisticated techniques for using the LLM to generate interview questions have not been studied in this work. For example, it is possible to create and fine-tune the LLM on an interviewer-interviewee conversation dataset, apply Reinforcement Learning from Human Feedback (RLHF) on interview quality using human preferences~\cite{OWJ+22}, use Retrieval-Augmented Generation (RAG) to automatically retrieve relevant and prior interview excerpts to inform follow-up questions~\cite{LPP+20}, use multi-agent approaches to create reflective loops wherein one LLM evaluates another LLM's questions~\cite{ZCS+23}, or apply domain-specific additional pretraining~\cite{XAA24}, if adapting the approach to relatively rare domains. We leave the exploration of such techniques to future work.

\section*{Data Availability Statement}
A replication package, including datasets and code, is available at Zenodo\footnote{https://zenodo.org/records/15793870}\cite{SSB+25}.

 \section*{Acknowledgment}
This work was supported by NSF Awards \#2217572 and \#2007298.

\end{document}